\documentclass{JHEP} 
\usepackage{amsmath}
\usepackage{epsfig}
\usepackage{amssymb,amsfonts}

\newcommand{\Tr}{{\rm Tr}\ }

\newcommand{\D}{{\cal D}}

\newcommand{\M}{{\cal M}}
\newcommand{\N}{{\cal N}}
\renewcommand{\S}{{\cal S}}

\newcommand{\Z}{{\cal Z}}
\newcommand{\pa}{\partial}

\newcommand{\Zint}{\mathbb{Z}}
\newcommand{\Real}{\mathbb{R}}
\newcommand{\Comp}{\mathbb{C}}
\newcommand{\bea}{\begin{eqnarray}}
\newcommand{\eea}{\end{eqnarray}}
\newcommand{\be}{\begin{equation}}
\newcommand{\ee}{\end{equation}}
\newcommand{\ir}[1]{\ensuremath{\boldsymbol{#1}}}
\renewcommand{\o}{\oplus}

\title{Nahm's equations, ${\cal N}=1^*$  domain walls, \\  and  D-strings in
$AdS_5\times S_5$}

\author{C. Bachas\\
 Laboratoire de Physique Th{\'e}orique
de l' Ecole Normale Sup{\'e}rieure\thanks{Unit{\'e} mixte  du
CNRS et de l'ENS,  
UMR 8549.} \\ 
24  rue Lhomond, {}F-75231 Paris Cedex 05, France\\
E-mail: \email{bachas@lpt.ens.fr}}

\author{J. Hoppe\\
Albert Einstein Institut, Max Planck Institut f{\"u}r Gravitationsphysik\\
Am M{\"u}hlenberg 1, D-14476 Golm, Germany\\
E-mail: \email{hoppe@aei-potsdam.mpg.de}}

\author{B. Pioline\footnote{On
leave of absence from LPTHE, Universit{\'e} Pierre et Marie Curie,
PARIS VI and Universit{\'e} Denis Diderot, PARIS VII, Bo{\^\i}te
126, Tour 16, 1$^{\it er}$ {\'e}tage, 4 place Jussieu, F-75252
Paris CEDEX 05, FRANCE}\\
Jefferson Physical Laboratory, Harvard University\\
Cambridge, MA 02138, USA\\
E-mail: \email{pioline@physics.harvard.edu}}

\abstract{We consider the following two  problems: classical 
domain walls in the ${\cal N} =1^*$  mass deformation of the maximally
supersymmetric Yang Mills theory, and D-strings as external magnetic
sources in the context of the AdS/CFT correspondence. We show that 
they are both described by Nahm's equations with unconventional
boundary conditions, and analyze the relevant moduli space of solutions.
We argue  that general  `fuzzy sphere'  configurations of
D-strings in AdS$_5$ 
correspond to Wilson-'t Hooft lines  in higher representations
of the dual $SU(n)$  gauge theory.}

\preprint{hep-th/0007067\\HUTP-00/A025\\LPTENS-00-30\\LPTHE-00-29}

\begin{document}

\section{Introduction, summary and directory}

The maximally supersymmetric (${\cal N}=4$ in four dimensions) 
Yang Mills theory has  an interesting  deformation that  
preserves ${\cal N}=1$  supersymmetry
while lifting all its flat directions \cite{Vafa:1994tf}. 
This theory (dubbed `${\cal N}=1^*$')  has a rich phase structure,
studied both  with  duality and instanton techniques 
\cite{Donagi:1996cf,S,Dorey:1999sj,Dorey:2000fc}, 
and more recently in the context of the
AdS/CFT correspondence \cite{Girardello:2000bd,Polchinski:2000uf,
Aharony:2000nt}. 
A particularly intriguing feature of the  theory is that it has a
large number of disconnected,  supersymmetric ground states. 
For a gauge group $SU(N)$ 
classical vacua  are labelled by the $N$-dimensional
representations $\rho$  of $SU(2)$. These  are determined 
uniquely by the dimensions of the irreducible blocks, i.e. by
a partition of $N$ in positive integers. The 
number of inequivalent vacua grows therefore 
as  $\sim e^{\sqrt{N}}$  when $N$ is large.

   One of the goals of the present work is 
to study the existence and  moduli space of classical
BPS domain walls interpolating between 
any two of these ground states.\footnote{For earlier work on BPS
walls in globally supersymmetric theories 
see \cite{Abraham:1991nz,Cvetic:1991vp,Cecotti:1993rm,
Dvali:1997xe,Witten:1997ep,Gibbons:1999np}.}
 The  problem
has been discussed in  the
 dual supergravity description of the ${\cal N}=1^*$
theory
in \cite{Dorey:2000fc,Polchinski:2000uf,Aharony:2000nt}, with 
a  particular emphasis on  walls  separating the Higgs from the 
(oblique) confining phases. Here we will consider the 
general situation but in the opposite, classical
gauge-theory limit.
Note that the  existence of domain walls interpolating
between a  pair of discrete  degenerate  vacua 
at infinity is not a priori
guaranteed  on topological grounds: it could , for instance, be energetically
favorable for an interpolating configuration to decompose into
a pair of mutually-repelling  walls with a different 
ground state in the middle.

  What renders our problem tractable is the observation that  classical
 BPS domain walls   are described by solutions to the much studied  Nahm
equations.  These  arise in a number of situations related
to monopole physics, and with a variety of boundary conditions  
\cite{Nahm,hitchin,Ganoulis:1982sx,Kronheimer}. 
The boundary conditions relevant for us  turn out to be
 precisely the ones considered previously by Kronheimer \cite{Kronheimer}
in his study of SO(3)-invariant antiself-dual connections 
on $S^3\times \Re \simeq S^4$. Building on his results we will 
bring out the following points: 
\begin{itemize}
\item[(a)] Supersymmetric walls exist
for each  pair $(\rho_-,\rho_+)$
of vacua such that the nilpotent orbit associated to the 
representation $\rho_+$ is contained in the closure of the nilpotent orbit
associated to $\rho_-$. This condition defines a {\it partial ordering}
($\rho_+<\rho_-$) between $SU(2)$ representations, which is 
easily read off from Young tableaux as illustrated
in Figure \ref{partord}.
In particular, for a BPS domain wall to exist,
it is necessary that the vacuum with the higher
superpotential does not contain  more irreducible blocks than the one with
lower superpotential, and that the  size of
the biggest block does not increase as the superpotential
 decreases. 
\item[(b)] The moduli space of domain walls is a singular hyperk{\"a}hler
manifold, whose dimension can be computed by Morse theory. 
 The dimension  satisfies an additivity rule displayed
in Equation  \eqref{addrule}.
\item[(c)]   This additivity of the number of massless modes
suggests that all BPS walls can be decomposed into 
some `elementary constituents' that 
interpolate between pairs of ordered  neighbouring ground states.
 These  elementary walls have no moduli other than
those dictated by the gauge  invariance   and global R-symmetry of the
problem.
\end{itemize}
These facts can be  verified readily in the tables provided in section 4,
which display the moduli spaces of domain walls 
 of the $SU(N)$ $\N=1^*$   theories, for all $N\leq 6$.
We emphasize that our study is  purely classical: it  uses the tree-level
 superpotential and  K{\"a}hler potential, and does not take into account
the splitting of certain classical vacua into several oblique-confinement
states. Which classical BPS walls are preserved at the
full quantum level, is thus an open  question that requires further
investigation, possibly 
along the lines of  \cite{Dvali:1997xe,Kaplunovsky:1999vt}.
 Note however that since
the dimension of the moduli space is given, as we will argue,
 by a Morse index, we expect it to be at least 
 robust under small deformations of the K{\"a}hler potential.

\vskip 0.4cm
One of the reasons for studying the ${\cal N}=1^*$
theory, apart from its possible relevance to pure QCD, is
its relation to the dielectric effect \cite{Myers:1999ps},
whereby D$p$-branes in a Ramond electric background 
expand into D$(p+2)$-branes with the topology of 
a `fuzzy' sphere. The gravity dual of the ${\cal N}=1^*$ theory
 exhibits  a supersymmetric version of this dielectric effect:
the Higgs/Coulomb vacua are  described by  `fuzzy' D5 branes 
embedded in the 
$AdS_5\times S^5$ geometry \cite{Polchinski:2000uf,Aharony:2000nt}
(for other manifestations
of this effect, see for example \cite{Kabat,DBS,sus,li,Ale,Jev}). 
{ The domain walls separating the  Higgs from confining vacua 
can be described in this picture by 
5-brane junctions \cite{Polchinski:2000uf}.
We expect a similar interpretation for the more general domain walls,
but we will not pursue this issue  here any further.}

We will,  instead, consider another stringy situation for which 
the same equations apply, namely the problem of D-strings
stretching out radially in $AdS_5\times S^5$. 
D-strings ending on separated D3-branes in flat spacetime are described
by the Nahm equations with standard boundary conditions -- this
gives a  concrete realization \cite{Dia} of the celebrated
ADHMN construction. What  we will show is that  D-strings in
the near-horizon geometry of the  D3-branes are controlled by the same
equations, but with boundary conditions identical to those of
Kronheimer's problem. This follows from the superconformal symmetry
of the worldvolume theory. The vacuum configurations of N
 D-strings in $AdS^5$ are thus labelled again  by 
N-dimensional representations $\rho$ of SU(2). 
These  supersymmetric `fuzzy-sphere' bound states
correspond in the holographic dual theory
to  heavy magnetic sources 
in non trivial representations of $SU(\infty)$.\footnote{
We thank J. Maldacena for an early suggestion of this point.}
The domain walls, which are the main theme of  this paper,  are
kinks  on the D-string worldsheet -- they
correspond to   braiding  operators  of the   Wilson-'t Hooft 
line in the dual CFT.

There is a number of related problems 
that we do not address.  Our matrix domain
walls also describe, for instance, supersymmetric instantons 
in the deformed matrix quantum mechanics 
of references \cite{Porrati:1998ej,Kac:2000av,Nekrasov:1999cg},
and may  bear on the vacuum structure of M(atrix) theory and on
scattering with longitudinal momentum transfer \cite{Pouliot,Hacquebord}.
They also occur as boundary RG flows between Cardy states of 
$SU(2)_k$ WZW models, much as in the Kondo problem.
It may also be interesting to find a string realization of
Kronheimer's original problem,
that of  $SO(3)$-invariant instantons in $S^4$. 
Finally,   the behaviour of
closed  (circular or rectangular) Wilson-'t Hooft 
lines in higher representations of the gauge group is a very
interesting problem that we do not address.

The organization of this paper is as follows. In section 2
we recall the vacuum structure of the $\N=1^*$ theory and
{ derive the equation as well as some obvious solutions for
the classical BPS domain walls.}
 In section 3, we discuss the relation
to the standard monopole problem, and in particular the limit
where a  non-abelian gauge symmetry is restaured.
Section 4 is an analysis of the
existence of solutions interpolating between  arbitrary 
representations, and relies a lot  on Kronheimer's work.
Section 5 considers the apparently unrelated problem of radial D-strings
in $AdS_5\times S^5$. We  show  that the transverse coordinates
of the D-string satisfy conditions isomorphic to the 
$\N=1^*$ domain walls, and  discuss the holographically 
dual interpretation of the solutions.
There is no harm in skipping the technicalities 
of section 4 in a first reading.
\enlargethispage{5mm}


\section{Supersymmetric domain walls in $\N=1^*$ SYM}
The $\N=4$ Yang-Mills theory in four  dimensions  admits a deformation
lifting all of its moduli space while preserving
$\N=1$  supersymmetry.
This deformation is most easily described by rewriting
the six real scalar fields  as three 
chiral superfields $\Phi^a$, taking their values 
in the complexified Lie algebra ${\cal G}^{\Comp}$ of the gauge group $G$.
One can then deform the $\N=4$ superpotential by a mass term, 
\begin{equation}
\label{spot}
W=\Tr \sum \left(\frac16~ \epsilon^{abc} \Phi^a [\Phi^b, \Phi^c]
-\frac12 m~ \Phi^a \Phi^a
\right)\ ,
\end{equation}
which breaks the   R-symmetry group from $SO(6)$ to $SO(3)$.  
We will use the same symbol for the  superfield and for its scalar
component, since  the context will make clear what we mean.
By redefining the phase of the fields, we can assume that $m$ is 
real and positive. The $D$-term contribution to the scalar potential, 
\begin{equation}
\label{dterm}
V= \frac{1}{g^2}
\Tr D^2\ \ \ {\rm with }\ \ \  D=\sum [\Phi^a, (\Phi^a)^\dag] \ .
\end{equation}
is not affected by the mass deformation.

\subsection{Vacua of mass deformed $\N=4$ Yang-Mills}

The supersymmetric classical vacua of the theory obey 
\begin{equation}
\frac{\pa W}{\pa \Phi^a}=  \frac12 \epsilon^{abc} [\Phi^b, \Phi^c]
-m \Phi^a  = 0 \ ,\quad D=0\ , 
\end{equation}
and are hence in one-to-one  correspondence with inequivalent 
 embeddings of $SU(2)$
in ${\cal G}^{\Comp}$.
For $G=SU(N)$ these are simply 
$N$-dimensional  representations, $\rho$, of $SU(2)$. 
Such representations can be always unitarized, meaning that the
three generators can be made antihermitean by a change of basis.  
The  $D$-term conditions force this  change of basis to be unitary, 
so that it can be undone  by a gauge transformation.
The vevs of the chiral fields in a vacuum $\rho$ are thus given by
\begin{equation}
\label{vacua}
\Phi^a= m \rho^a\ ,\quad  [\rho^a,\rho ^b]=
 \epsilon^{abc}\rho^c\ ,\quad \rho^a=-(\rho^a)^\dag
\end{equation}

The representation $\rho$ is in general reducible, and can be
decomposed into different blocks of (integer or half-integer) spin $j$,
\begin{equation}
\label{sumspin}
\rho\;=\;\bigoplus\; n_{j}\;[j]\ \ \ \ \ {\rm with} \ \ \
\quad \sum (2j+1)\;n_{j}=N\ .
\end{equation}
The unbroken gauge symmetry in the corresponding vacuum   
is 
\begin{equation} 
G_{(\rho)}\equiv   \left( \prod_j  U(n_j) \right) \Bigl/ U(1)\ .
\end{equation}
 This contains in  general
abelian factors, which are  asymptotically-free  in the infrared. 
Vacua with a mass gap (and no abelian factors) correspond to  the special
representations which  break up 
precisely  into $k$ identical blocks
of size $N/k$. There is one such representation for each divisor $k$ of $N$.
The corresponding  vacuum   splits  at the quantum level
into $k$  Higgs, confining and oblique vacua,  related by the
spontaneously broken R-symmetry.

For $G=SO(N)$ (or  $USp(N)$), the embeddings of $SU(2)$ are again 
given by $N$-dimensional representations which must now
 be chosen real (respectively  pseudo real). This implies that only 
{ integer} (respectively  half-integer) spins  appear in the decomposition
\eqref{sumspin}. Unbroken  $U(1)$ gauge  symmetries  now arise 
whenever there is a pair  of identical representations in the
decomposition, i.e. for every $j$ for which   $n_j=2$. 
As for embeddings of $SU(2)$ in the exceptional Lie groups, these
have been classified in reference \cite{Dynkin}.

\subsection{Supersymmetric domain walls}
A theory with many  isolated vacua is guaranteed, under
some mild assumptions, to have smooth domain-wall solutions.
For vacua with unbroken supersymmetry these walls may,  but need
not a priori,  be supersymmetric. 
 To see why  let us review 
the argument leading to a  BPS bound on
the tension of  domain walls   in 
$\N=1$ supersymmetric theories in four dimensions
(see for instance \cite{Gibbons:1999np}). 
The starting point is  the energy functional for static
 configurations of the chiral fields,  
\begin{equation}
E = \int \; \frac{d^3r}{g^2}~ \Tr \left(
\left|\nabla \Phi^a\right|^2 + 
\left|\frac{\pa W}{\pa \Phi^a}\right|^2 + D^2 \right)\ .
\end{equation}
Since we are interested in 
planar,  $SO(1,2)$-invariant domain walls, we have set 
 the gauge fields equal to zero. 
This is consistent as long as  the Gauss conditions
are  satisfied, 
\begin{equation}
 \sum_a\  \left[ \Phi^a, (\nabla\Phi^a)^\dag \right] +
\left[ (\Phi^a)^\dag, \nabla\Phi^a \right]  = 0\ .
\label{gauss}
\end{equation} 
The scalar fields are, furthermore, functions only of $x$,
which is the
  coordinate parametrizing  the transverse direction.
The tension of the wall ($T\equiv  E/{\rm Area}$)   can 
then  be written  as a sum of squares plus  a boundary term,  
\begin{equation}
T  =\int_{-\infty}^{+\infty}     \frac{dx}{g^2}~ \left(
\Tr \left| \left( {d{\Phi}^a\over dx}\right)^\dag
 - e^{i\alpha} \frac{\pa W}{\pa \Phi^a} \right|^2
+ 2\frac{d}{dx}\Re ( e^{i\alpha}W ) + \Tr D^2 \right)\ . 
\end{equation}
This expression leads  to the lower (BPS) bound
\begin{equation}
T  \;\geq \; {\rm sup}_\alpha \; \frac{2}{g^2}~ \Re ( e^{i\alpha}W
)\Bigr\vert_{-\infty}^{+\infty}\; =\; {2\over g^2}~  |\Delta W|  \ , 
\end{equation}
where $\alpha$ is  any  constant phase, and $\Re (A)$
stands for the real part of the quantity $A$.

The strictest  bound is obtained when  $e^{-i\alpha}=\Delta W/|\Delta W|$,
and it is  saturated by  solutions of  the
first-order equations
\begin{equation}
\left({{d\Phi}^a\over dx}\right)^\dag  
 = e^{i\alpha} \frac{\pa W}{\pa \Phi^a}\ , 
\label{bps}
\end{equation}
provided  the $D$-terms, and the Gauss constraints
also vanish.
Note that the above equations imply
\begin{equation}
\label{walpha}
\frac{d}{dx} W(\Phi) = e^{i\alpha} \left| \frac{\pa W}{\pa \Phi^a}
\right|^2\ ,
\end{equation}
so that the superpotential `moves'  along 
a straight line, in the direction $\alpha$,  on the complex plane.
Since in  our case  the superpotential at all the  vacua is real,
we may choose $e^{i\alpha}=-1$ 
(the choice $e^{i\alpha}=1$
corresponds to flipping the sign of  the coordinate $x$, which exchanges
 walls  and  anti-walls).
The  superpotential must  thus  be a real
 decreasing function from left to right.
The BPS conditions \eqref{bps}
can, in fact, be interpreted as the equations of {\it gradient flow}
for the potential $\Re (W)$. For a  BPS-saturated 
wall to  exist,  gradient flow between two
critical points of $\Re (W)$  must be  allowed.

The $\N=1^*$ superpotential evaluated at the vacuum $\rho$ is 
proportional to the trace of the quadratic Casimir,
\begin{equation}
W{(\rho)} = m^3 \sum_j  \frac{j(j+1)(2j+1)}{6}  \;n_{j} \ .
\label{values}
\end{equation}
For more general gauge groups $G$,   the vacuum values  of $W$ 
are equal to  $m^3/4$ times the Dynkin indices, $d(\rho)$,
of the corresponding $SU(2)$ 
representations.  A complete list of the $d(\rho)$
 can be found in reference \cite{Dynkin}.
According to our previous discussion, BPS 
domain walls interpolating between
$\rho_-$  and $\rho_+$, 
\begin{equation}
\Phi^a(x\to-\infty)\to m {\rho_-^a}\ \qquad {\rm and} \qquad
\Phi^a(x\to+\infty)\to m {\rho_+^a}\ ,
\label{tbound}
\end{equation}
may  exist  only if  $W(\rho_-)>W(\rho_+)$. 
Anti BPS domain walls interpolate, of course, in the
opposite direction. Walls separating two vacua with
$W(\rho_-)= W(\rho_+)$ are, on the other hand, 
necessarily {\it non-supersymmetric}. Such stable non BPS branes are
generic when $N$ is large, since there are  exponentially many vacua 
and only polynomially many possible  values for \eqref{values}.  
Quantum corrections may lift this large degeneracy, but we will not
pursue this question here further.

\subsection{Some explicit  solutions}

A simple way of satisfying the $D$-term constraints \eqref{dterm} is
by restricting the chiral fields $\Phi^a$  to be {\it antihermitean}.
This is also the condition that would arise if we considered
four-dimensional $N=1$ Yang-Mills reduced to $D=1$.
The BPS equations then read
\begin{equation}
\frac{d\Phi^a}{dx} =  \frac12 \epsilon^{abc} [\Phi^b, \Phi^c ]
-m \Phi^a\ .
\label{nahmt}
\end{equation}
As  can be verified easily 
the Gauss conditions are then also
automatically obeyed. Supersymmetric domain walls will thus
exist  whenever the  equations \eqref{nahmt} admit  
antihermitean solutions. The converse 
need  not  be, a priori, true -- 
we will  discuss  this question later 
in section \ref{Morse}.

Some simple solutions of the above equations can be readily found.
Using for instance the ansatz~   $\Phi^a = m f(mx)~ \rho_-^a$ , 
leads to the differential equation~
$ f^{\ \prime}  =f(f-1)$. This can be integrated to give 
\begin{equation}
\label{expsol}
\Phi^a(x)=\frac{m}{1+e^{m(x-x_0)}}~ \rho_-^a\ ,
\end{equation}
which is  a domain wall  
interpolating  between any initial  vacuum  $\rho_-$ and the 
vacuum with unbroken gauge symmetry, 
 $\rho_+ = [0]\oplus\cdots \oplus [0]
$\  ($N$ times). A slight  modification of this ansatz
leads to solutions that interpolate between
$\rho_-= \rho\otimes {\tilde \rho}$
 and $\rho_+ = {\tilde \rho}\oplus \cdots \oplus {\tilde \rho}$\ 
($\dim \rho$\  times), for any pair
 $\rho$ and ${\tilde \rho}$ of representations.
 An  explicit wall profile in this case is
\begin{equation}
\Phi^a(x)= \frac{m}{1+e^{m(x-x_0)}}~ \rho^a \otimes 1 +
 m~1 \otimes {{\tilde \rho}}^{\ a} \ .
\end{equation}
It can be checked that this solves the BPS equations, and obeys 
the appropriate boundary conditions. A particular example of this type is
the wall that interpolates 
$[j-1/2]\oplus [j+1/2]\to [j]\oplus [j]$.

 We do not know of a 
systematic method to
construct explicit solutions of \eqref{nahmt} in general.
We  will be able, nevertheless,  to  characterize their moduli spaces 
in section 4.

\section{Nahm's equations and non-abelian monopoles}

The form of \eqref{nahmt} is  reminiscent
of the much studied Nahm equations, which give a dual description
of $SU(2)$ monopoles on $\Real^3$ \cite{Nahm,hitchin}.
The only difference is the mass term, but this can be eliminated by
the change of variables 
\begin{equation}
 \Phi^a \equiv -e^{-mx}X^a \qquad  {\rm and} \qquad s \equiv e^{-mx}/m\  ,
\end{equation}
which brings  
\eqref{nahmt}  to  the standard Nahm form, 
\begin{equation}
\frac{dX^a}{ds} = \frac12 \epsilon^{abc} [X^b,X^c] \ .
\label{nahmu}
\end{equation}
Note that 
$s$  takes values  on the semi-infinite interval $]0,+\infty[$ ,
and the boundary conditions at both ends are
\begin{equation}
\label{ubound}
X^a(s\to 0) \sim -\frac{\rho_{+}^a}{s} + {\rm finite}\ ,\quad
X^a(s\to \infty)
 \sim -\frac{\rho_-^{a}}{s} + {\rm subleading}\ .\quad
\end{equation}
These differ from  the boundary conditions in Nahm's
description of the standard $N$-monopole problem \cite{Nahm,hitchin}, 
where  the $X^a$ must  have
 poles  at both ends of a {\it finite} interval,
with residues  given by the {\it same} {\it irreducible} 
representation $\rho$ of dimension $N$.
In this section  
 we would like to 
discuss this issue further.

Nahm's construction arises very naturally 
in a type IIB string-theory setting, where the monopoles 
are D-strings stretching  between parallel D3-branes. A $N$-monopole
corresponds to  $N$ D-strings whose coordinates are hermitean $N\times N$
matrices. What the  Nahm equations describe is
the evolution of these  transverse matrix coordinates, 
  as one moves along the 
D-string worldsheet \cite{Dia,KST,Brecher}.
 We may  visualize the  D-strings as forming  a `fuzzy'
spherical D3-brane,  whose radius blows up at both  ends
 of the finite $s$-interval, where the transverse coordinates have a pole. 
These blown up D-strings  are in fact 
indistinguishable from  the  D3-branes on which they terminate
\cite{CMT}, just as a  D-string has a  dual description  as a spike
or `BIon' of the D3-brane \cite{mald,Gibbons:1999te}.
One may therefore forget the D3-branes  altogether,
and simply ask for a pole $X^a \sim \rho^a/(s-s_0)$ 
 at the desired locations.


\EPSFIGURE[t]{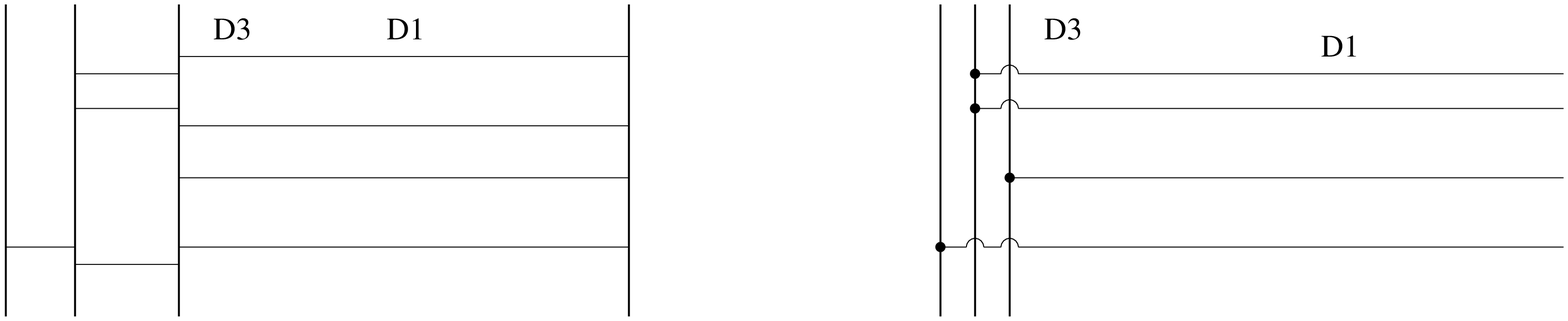,width=15cm}{Left: D1/D3-brane
 configuration 
describing  a ($N_1$,$N_2$,$N_3$) = (1,3,4)   mutlimonopole  of  $SU(4)$.
 Right: D-brane configuration that could correspond to  Nahm's equations
on the semi-infinite line,  with a reducible residue
 $\rho= [{1/2}]{\o}[{0}]{\o}[{0}]$
at the origin. The three D3-branes lie on top of each other.\label{d1d3}}


In the simplest case of  $G= SU(2)\to U(1)$  there are precisely two
poles, with 
residues controlled by the same irreducible 
representation $\rho$. This is
consistent  with the fact that the gauge theory can be  engineered with
 just two D3-branes.
In the  general case  $G= SU(n)\to U(1)^{n-1}$ the boundary
conditions are more  subtle  \cite{Hurtubise}, but are again easy
to visualize from the D-brane perspective \cite{KST}. 
The novel feature is
that the  D-strings can now both {\it intersect and  terminate} on the
D3-branes (see Figure 1). 
What one is instructed in the end to do is to  solve $(n-1)$ 
Nahm  equations for  matrices of size $N_i\times N_i$,
corresponding to $N_i$ D-strings  in the
$i$th interval. The $N_i$ are related in an obvious way to the monopole charges.
The $i$th and $(i+1)$ solutions must be glued together 
by requiring a 
pole on the $p \times p$ block
 of the larger matrices (where $p=|N_i-N_{i+1}|$),
and a step-function discontinuity on the remaining parts. 
This latter discontinuity is  controlled by the open strings that stretch
between  the D-strings on each side of the D3-branes \cite{KST}.

   There are two limits in which the above  description degenerates~: 
(i) a  D3-brane can move off to infinity, in which case some 
of the  monopoles  become infinitely heavy, or  (ii) two or more
D3-branes may coincide, in which case a non-abelian gauge symmetry is restaured and
some of the monopoles
become,  formally at least, massless.
 Moving a D3-brane to infinity makes one of
the $s$ intervals semi-infinite. Moving, on the other hand,
  two or more D3-branes
on top of each other forces  two or more poles (corresponding,  in
general, to  distinct irreducible  representations of $SU(2)$)
 to collide. One is thus tempted, at first sight, to conclude that 
\eqref{ubound} are the   
appropriate boundary conditions 
in these limits.

 The situation is,  however,  much more subtle. In a theory with a
 decoupled $U(1)$  the natural condition at infinity is that the
 $X^a$ approach   constant diagonal  matrices, whose entries  are the 
classical  positions of the singular monopoles (see
 \cite{Cherkis} for a detailed discussion).  Our boundary
 condition is different: the $X^a$  do approach 
 zero at infinity,  but  it is crucial that  the
 subleading behaviour be  specified.
 As we will argue  later,  in section 5,
 this boundary condition arises naturally
 in the curved  near-horizon geometry of D3-branes.

 What about the boundary condition at $s\to 0$~? If the  residue $\rho$
corresponds   to a  reducible representation of $SU(2)$ containing
$\tilde n=\sum n_j$ irreducible blocks, then 
we are  describing  monopoles in a point of
enhanced $SU(n\ge \tilde n)$ gauge symmetry. 
 This is indeed  consistent  with the fact that  
the D-strings grow into $\tilde n$ spherical D3-branes, whose   sizes all  diverge
simultaneously, 
as illustrated in figure 1 (there can of course also exist extra D3-branes
at this point).
Whether one can assign magnetic monopoles to representations of an unbroken
non-abelian gauge group has been, for many years,
 a subject of debate.\footnote{We thank T. Tomaras for bringing this
issue to our attention.
For a recent  discussion of the problem
and for earlier references see
\cite{Weinberg} .} 
We dont have much to add to  this debate here.
We  just note that there does exist  a natural correspondence
between residues  $\rho$ in the Nahm problem, 
and representations $r(\rho)$ of the  unbroken
gauge group. 
The Young tableau of the representation $r(\rho)$ has 
 a row of $(2j+1)$ boxes for each spin $j$ in the 
decomposition \eqref{sumspin}.
This is  illustrated in  Figure \ref{young}.


\EPSFIGURE[t]{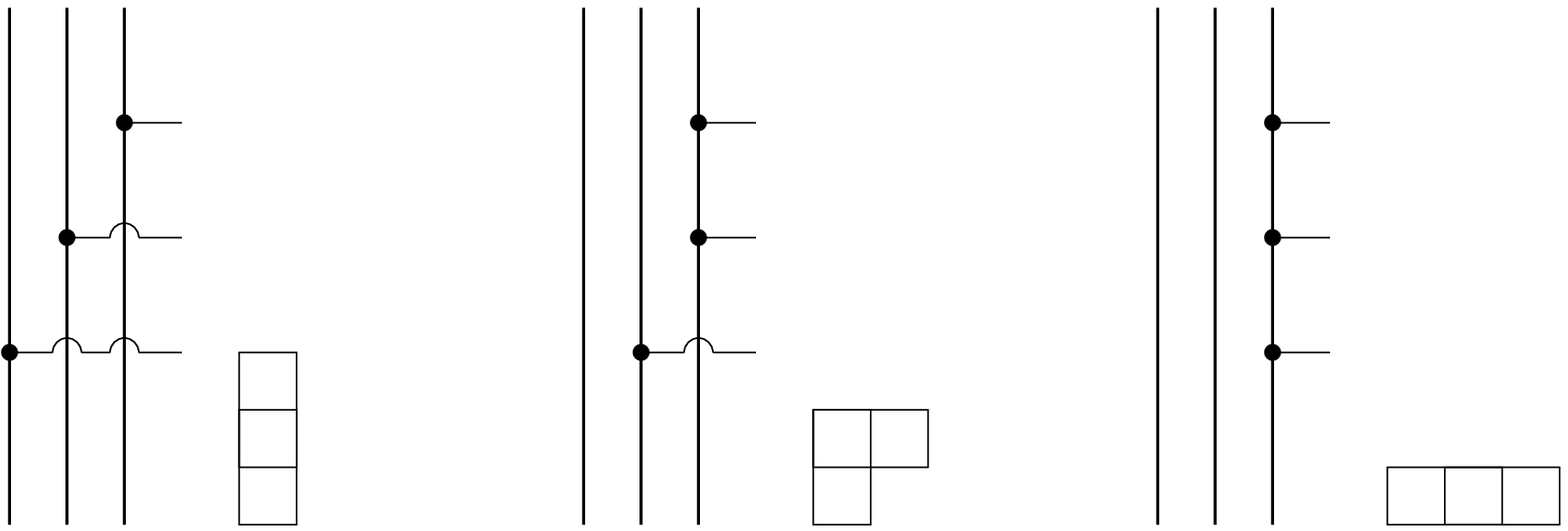,height=4cm}{D1-branes configurations and
associated Young tableaux as discussed in the text. \label{young}}


The above  assignement of a representation of the unbroken gauge group 
 passes two
simple consistency  checks: (a) the number of boxes in the Young tableau
 is the same as the number $N$
of D-strings -- this is consistent with the expected $n$-ality  
of the representation; (b) the number $\tilde n$ of rows
is the 
minimal number of Chan Paton charges required 
to build a state of the representation $r(\rho)$, because of
antisymmetrization. This fits nicely
with the fact that $\tilde n$  is  also the lower bound on the  number of
coincident D3 branes. 
We stress,  however,  that we have no clear  reason to assign complete
$SU(n)$ representations to our  Nahm problem  at present.
We will come back to this question in the AdS/CFT context of section 5.
The Young tableau of $r(\rho)$ is,
in any case, a useful graphical representation
of the ${\cal N}^*=1$ vacuum corresponding to $\rho$, as will become
apparent in section 4.3.


\section{The moduli space of domain walls}

{ Many of the techniques used
in the analysis of the standard monopole problem can be carried over to the
domain wall problem \eqref{nahmt}, or its mathematical equivalent
\eqref{nahmu} and \eqref{ubound}
 which describes, as we will see later,  D-strings in 
$AdS_5$. The unusual boundary conditions
introduce however some  new twists: 
in particular, the problem still admits a Lax
pair -- but the spectral curve is degenerate, and the moduli space 
is still hyperk{\"a}hler -- but singular.}

We are interested in  solutions to \eqref{nahmt}
interpolating between two different vacua, characterized by 
two arbitrary unitary representations, 
$\rho_-$ and $ \rho_+$ of $SU(2)$. The  embedding of
these representations inside the gauge group $G$ can be arbitrary.
To render this explicit we thus write the boundary conditions as:
\begin{equation}
\label{xbound}
\lim_{x\to -\infty} \Phi^a  = g~\rho_-^a~  g^{-1}\ ,\qquad \ {\rm and}\ \ \ \
\lim_{x\to +\infty} \Phi^a  = \rho_+^a \ , 
\end{equation}
where we fixed the embedding  at $+\infty$, and allowed
arbitrary rotations $g\in G$ at $-\infty$.
Remarkably, this very problem has been studied 
by Kronheimer \cite{Kronheimer}\footnote{We are grateful to 
O. Biquard and N. Hitchin for drawing our attention to this work.},
 in the context of
$SO(3)$-invariant anti-self-dual connections on $S^4$: 
the superpotential \eqref{spot} can indeed be thought as the Chern-Simons 
invariant of a left-invariant G-connection on $S^3$ specified by the three
matrices $\Phi^a$, and gradient flows of this functional yield
anti-selfdual configurations on $\Real\times S^3 \sim S^4$.
We shall now review Kronheimer's results, perform
some explicit computations of moduli spaces for particular choices
of the pair $(\rho_-,\rho_+)$, and draw conclusions as to the existence
of BPS domain walls.

\subsection{Lax pair, Moment map and Hyperk{\"a}hler structure} 
In analogy with the usual Nahm problem, we can  define 
in terms of the original variables $\Phi^a$ the matrices 
\bea
L&=&(\Phi^3+i \Phi^2)+2 i \zeta \Phi^1 + \zeta^2(\Phi^3-i \Phi^2)\\
M&=&i \Phi^1+\zeta(\Phi^3-i \Phi^2)\ ,
\eea
where $\zeta$ is the spectral parameter. 
Equation \eqref{nahmt} can now be written as
\begin{equation}
e^{-i\alpha}\zeta^2 \frac{\pa}{\pa x}{\bar L}(-1/\bar\zeta)=
[L(\zeta),M(\zeta)] - m L(\zeta) \ .
\end{equation}
Imposing   the reality condition
$\zeta^2 [L(-1/\bar\zeta)]^\dagger=-L(\zeta)$, 
which corresponds to  antihermitian matrices $\Phi$, we find
\begin{equation}
\frac{\pa}{\pa x} L(x,\zeta)=[L,M]- m L\ .
\end{equation}
This implies, in particular, the following   first-order linear
differential  equations:
\begin{equation}
\frac{d}{dx} \Tr L^n=- m n\;  \Tr L^n
\end{equation}
for all $n>0$.  In order for the matrices to remain finite at $\pm \infty$,
this requires $\Tr L^n=0$ for all $n>0$, or equivalently $L^N=0$.
This implies that the spectral curve $\det(L(\zeta)-\eta)=0$ degenerates 
to $\eta^N=0$. Our deformed
Nahm problem can therefore be thought as a singular limit of the usual 
monopole problem when all constants of motion vanish. 
In particular,
this implies that explicit solutions should be given in terms
of hyperbolic functions, instead of the elliptic functions arising
in the monopole problem. Indeed we have seen such an example
in section 2.3, and it would be interested to generalize it
to arbitrary pairs of vacua. Instead we shall study the moduli
space of such domain walls, using techniques similar as in 
the standard monopole problem \cite{Donaldson}.

In order to make the hyperk{\"a}hler
structure manifest, let us introduce a gauge field $\Phi^0$. 
The equation \eqref{nahmt} becomes 
\begin{equation}
\label{nahmt0}
\frac{d \Phi^a}{dx}+[\Phi^0,\Phi^a] = \frac12 \epsilon^{abc}
[\Phi^b,\Phi^c] - m \Phi^a
\end{equation}
and is invariant under the gauge transformations
\begin{equation}
\Phi^0 \to g\Phi^0 g^{-1}-\frac{dg}{dx}g^{-1}\ ,\quad
\Phi^a \to g \Phi^a g^{-1}
\end{equation}
where $g$ is an element of the (real) group $G$.
The Nahm equations \eqref{nahmt0} can now be interpreted as the three
moment maps for the action of the gauge group on the quaternionic
vector space of matrices $(\Phi^0(x),\Phi^1(x),$ $\Phi^2(x),\Phi^3(x))$ 
with metric
\begin{equation}
ds^2 = \int_0^{\infty}dx\; e^{mx}\;  \Tr[(\Phi^0)^2+(\Phi^1)^2+(\Phi^2)^2
+(\Phi^3)^2]\ .
\end{equation}
As usual, these equations can be split into a  complex and a  real one.
Defining  the complex variables
\begin{equation}
\alpha=\frac12 (\Phi^0-i \Phi^1)\ ,\qquad
\beta=-\frac12 (\Phi^2+i \Phi^3)\ ,\qquad
\end{equation}
we can express the equations \eqref{nahmt0}  as follows:
\begin{eqnarray}
\frac{d\beta}{dx} +m\beta + 2[\alpha,\beta]=0  \label{cxe}\\
\frac{d}{dx}(\alpha+\alpha^\dagger) + m(\alpha+\alpha^\dagger)
+ 2([\alpha,\alpha^\dagger]+[\beta,\beta^\dagger])=0 \ .
\label{ree}
\end{eqnarray}
The complex equation is now invariant under complex gauge transformations,
\begin{equation}
\alpha\to g\alpha g^{-1}- \frac12 \frac{dg}{dx}g^{-1}\ ,\quad
\beta\to g \beta g^{-1}\ .
\end{equation}
Following the original approach of Donaldson \cite{Donaldson},
we can split the problem of solving the Nahm equations in  two parts: 
(i) find all solutions of the complex equation 
with the required boundary conditions, modulo complex gauge transformations, 
and (ii) show that there exists 
a solution of the real equation in the complex conjugacy
class of each solution of (i).
Part (i) is purely topological, since
all solutions to the complex equation \eqref{cxe} are locally pure gauge.
Part (ii) can be proved to hold using variational 
techniques \cite{Kronheimer}. 
In the following we shall restrict ourselves to part (i), which
already yields the moduli space with a particular choice of complex
structure.

\subsection{Nilpotent orbits and the moduli space of domain walls\label{nil}}
Let us define
\begin{equation}
H_\pm =\rho_\pm \left( \begin{array}{cc} 1 &0\\0&-1 \end{array}\right)\ ,\quad
X_\pm =\rho_\pm \left( \begin{array}{cc} 0 &1\\0&0 \end{array}\right)\ ,\quad
Y_\pm =\rho_\pm \left( \begin{array}{cc} 0 &0\\1&0 \end{array}\right)\
. \quad
\end{equation}
It may then be shown \cite{Kronheimer} 
that any solution of the complex equation
\eqref{cxe} can be expressed  as follows:
\begin{eqnarray}
x\in ]-\infty,0]:&& \left\{ \begin{array}{ccc} 
\alpha_1(x)&=&  \frac{m}{4} H_- \\ 
\beta_1(x)&=&  \frac{m}{2} Y_- \end{array} \right.\\
x\in [0,\infty[:&& \left\{ \begin{array}{cccl} 
\alpha_2(x)&=&  \frac{m}{4} H_+ &\\ 
\beta_2(x)&=& \frac{m}{2}  Y_+ &+ e^{-mx}~e^{-m H_+ x/2} \cdot Z \cdot e^{m H_+ x/2}\ , 
\end{array}\right. 
\end{eqnarray}
modulo  a complex gauge transformation $g_-$ 
(respectively  $g_+$)  approaching a constant (respectively the identity) 
when  $x$ goes to $ - (+) \infty $~. 
Here $Z\in \Z(\rho_+)$ is  the centralizer of $X_+$ in $su(N)$.
The essential idea behind this claim is that any $\alpha$
approaching $H_\pm $ at $\pm \infty$ is gauge equivalent to the
constant $H_\pm $. The value of $\beta$ is then obtained by
solving the complex equation, and further adjusting the
gauge transformation to come as close as possible to $\beta=m Y_\pm /2 $.
Matching  the two solutions $(\alpha_1,\beta_1)$ and $(\alpha_2,\beta_2)$
 at $t=0$
we find that 
\begin{equation}
\frac{m}{2} Y_+ + Z \in \N(\rho_-) \cap {\S}(\rho_+)\ ,
\end{equation}
where $\N(\rho)$ denotes the set of elements of ${\cal G}^\Comp$
related to $Y(\rho)$ by conjugation under the complexified 
gauge group $G^{\Comp}$,
and ${\S}(\rho)$ is the affine space $Y(\rho)+\Z(\rho)$. The element
$z$ so constructed is also unique, so that we arrive at  Kronheimer's
result: {\it the moduli space
of solutions to Nahm's equations interpolating between $SU(2)$ representations
$\rho_-$ and $\rho_+$ is given by the intersection}
\begin{equation}
\label{modsp}
\M(\rho_-,\rho_+)= \N(\rho_-) \cap {\S}(\rho_+)\ .
\end{equation}
In order to appreciate the significance of this result, it is useful
to recall a number of properties of the spaces $\N(\rho)$ and
${\S}(\rho)$ \cite{Slodowy,Kronheimer}:

(i) There is a one-to-one correspondence between the conjugacy classes
of embeddings of $SU(2)$ into ${\cal G}$ and the $G^\Comp$-orbits
of nilpotent elements in ${\cal G}^\Comp$. For $G=SU(N)$, this
correspondence is provided by the Jordan canonical form.
This justifies the notation $\N(\rho)$.

(ii) ${\S}(\rho)$ is a slice of ${\cal G}^\Comp$ transverse to the
nilpotent orbit $\N(\rho)$, which it intersects only at the
origin $Y$. In particular, for $\rho_-=\rho_+$, the moduli
space reduces to a point $Y$, which describes the vacuum
(no domain wall).

(iii) ${\S}(\rho_+)$ is the solution to a linear problem, namely
the centralizer of the matrix $X_+$. On the other hand, $\N(\rho_-)$
is given by solutions to a set of polynomial equations
( among which $s^n=0$, where $n(\rho_-)$ 
is the order of nilpotency of $\rho_-$). The moduli space
is therefore a $\rho_-$-dependent 
algebraic variety in the space ${\S}(\rho_+)$.

(iv) For $\rho_+=\ir{1}^N$ the trivial representation, we have
${\S}(\rho_+)=G^{\Comp}$, so that $\M(\rho_-,\rho_+)= \N(\rho_-)$: hence
any representation $\rho_-$ can be interpolated to the trivial
representation through a solution of Nahm's equations. Indeed we 
have found an explicit example of such a solution in \eqref{expsol},
but there is in fact a moduli space $\N(\rho_-)$ of them.

As an example of this construction, 
let us consider the simplest case of the domain
wall interpolating between the irreducible representation $\rho_-=[N]$ 
of $SU(2)$ of dimension $N$ (or more generally, the {\it principal}
or {\it regular} embedding of $SU(2)$ into $G$) and $\rho_+=[N-1]
{\o}1$ the {\it subregular} embedding. Since the subregular orbit 
has complex codimension 2 in the regular nilpotent orbit, the intersection
$\N(\rho_+)\cap{\S}(\rho_-)$ has complex dimension 2. 
For $G$ simply-laced, the moduli
space is in fact \cite{Kronheimer} 
the ALE space $\Comp^2/\Gamma$, where $\Gamma$
is the discrete subgroup of $SU(2)$ of the same ADE type as the
gauge group $G$. This is easily checked by explicit computation in the
$G=SU(N)$ case: the matrices in ${\S}(\rho_-)$ form a
$(N+2)$-dimensional subspace of $\Comp^{N^2}$ parameterized 
by\footnote{For the purpose of computing
the dimension, we work  in an non-normed basis, taking $X$ in the Jordan 
form with coefficients 1 above the diagonal.}
\begin{equation}
s=\left( \begin{array}{ccccc|c}
a_1 & a_2 & a_3 & \cdots & a_{N-1}  & b\\
1   & a_1 & a_2 & \cdots & a_{N-2}  & 0\\
0   &  1  & a_1 & \ddots        & \vdots  &\vdots\\
\vdots &\ddots  & \ddots & \ddots & a_2 & \vdots \\
0      & \cdots  &   0     & 1    & a_{1} & 0\\ \hline
0      & \cdots  & \cdots   & 0     & c       & d
\end{array}\right)\ .
\end{equation}
where all but one of the entries on the lower diagonal are fixed
and we set $m=2$.
Requiring $z\in \N(\rho_+)$ is equivalent to imposing $s^N=0$.
This imposes the tracelessness condition $d=-(N-1)a_1$
and relates the off-diagonal values $a_i=\alpha_i a_1^i$
to the diagonal coefficient $a\equiv a_1$, where $\alpha_i$ are computable
numerical coefficients. Finally, it imposes $bc= \alpha_N a^N$,
which we recognize as the complex equation of the $A_{N-1}$ singularity.
This space is the singular limit
of the well-known 4-dimensional gravitational instantons \cite{Kron2}.
It would be interesting to understand under what circumstances
the singularity might be resolved.

\begin{figure}
\centerline{\mbox{\epsfysize 3cm \epsffile{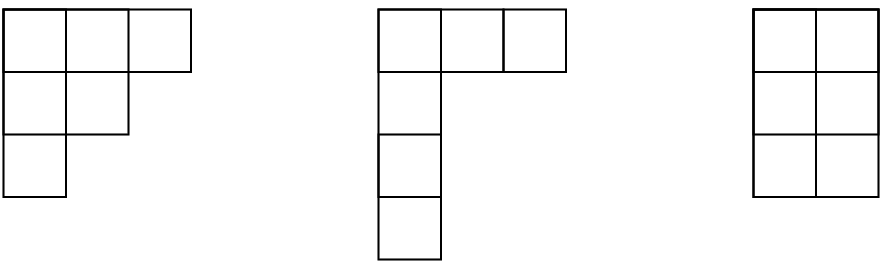}
\put(-290,-25){$\ir{3}\o\ir{2}\o\ir{1}$}
\put(-170,-25){$\ir{3}\o\ir{1}\o\ir{1}\o\ir{1}$}
\put(-45,-25){$\ir{2}\o\ir{2}\o\ir{2}$}
\put(-210,30){$<$}
\put(-80,30){$\not >$}}}
\caption{Partial order between $SU(2)$ representation. The number
of squares contained in the first $p$ columns of the Young tableau has to
increase for all $p$. The two representations on the right cannot
be related by a BPS nor an anti-BPS domain wall.\label{partord}}
\end{figure}

This strategy of parameterizing the linear space ${\S}(\rho_+)$ and
imposing the nilpotency of $z$ can be applied to any choice of
representations $(\rho_-,\rho_+)$, although it soon becomes 
rather cumbersome. Yet there is a simple way of deciding whether
a BPS domain wall exists or not. Indeed, if such a wall exists,
it will at least have a translational zero-mode, so that the
moduli space $\N(\rho_-)\cap {\S}(\rho_+)$ has to be non-empty.
This condition is in fact equivalent to
\begin{equation}
\label{clos}
\N(\rho_+)\subset\overline{\N(\rho_-)} \ .
\end{equation}
This gives a necessary and sufficient condition for the existence
of BPS domain walls interpolating between two representations. 
This condition can be further explicited by noting
that (for $SU(N)$) a nilpotent orbit (or a representation $\rho$) is uniquely
labelled by the vector $k_p(\rho), p=1\dots N$ where 
$k_p(\rho)$ is the dimension of the null space of $s^p$, or equivalently
the number of blocks in the Jordan decomposition of $s^p$. In particular,
$k_1(\rho)$ is the number of irreducible blocks appearing in $\rho$, 
$k_N=N$ by definition, and
\begin{equation}
\label{clos2}
k_p(\rho)= 2 n_{2} + 3 n_3 + 4 n_4 + \dots + (p-1) n_{p-1}
     + p ( n_p + n_{p+1} + \dots + n_N )
\end{equation}
where the $n_{2j+1}$ are the number of irreps of dimension $2j+1$ appearing
in $\rho$, as in \eqref{sumspin}. All the $k_p(\rho)$ can only
stay constant or increase as one goes from $\N(\rho)$ to its
closure, hence the condition \eqref{clos} can be rewritten as
\begin{equation}
\label{cns}
k_p(\rho_-) \leq  k_p(\rho_+)\ ,\quad p=1\dots N
\end{equation}
This condition can be understood graphically in terms of Young tableaux
associated to each reducible representation of $SU(2)$ as in Figure
\ref{partord}: $k_p(\rho)$ is the number of squares appearing in
the first $p$ columns of the Young tableau, and has to increase
from $\rho_-$ to $\rho_+$. 
In particular, the number of irreps (or fuzzy D$(p+2)$-branes in
a more physical language) can only increase from $-\infty$ to
$+\infty$, in addition to the decrease of the superpotential. 
It is also important to note that the order 
between $SU(2)$ representation is only partial, and there exists pairs of vacua
with no BPS domain wall interpolating between them. This provides
an interesting example of spontaneous supersymmetry breaking
at a classical level.

\subsection{Additivity rule and elementary domain walls}
Let us start by discussing the structure of the affine space ${\S}(\rho)$,
in the case of $G=SU(N)$. Taking for $\rho$ the irreducible representation
of dimension $N$, it is easy to see that the centralizer $Z(\rho)$ has
dimension $N$: it consists of upper triangular matrices, with coefficients
equal along the upper diagonals. If $\rho=\oplus_i \rho_i$
is reducible, the structure is similar in each $\dim(\rho_i)\times 
\dim(\rho_j)$
block, with $\min[\dim(\rho_i),\dim(\rho_j)]$ free coefficients in each block.
Ordering the representations $\rho_i$ by decreasing dimension
and counting them with multiplicity\footnote{Note that the labelling
of representations here is different from the one used in \eqref{sumspin}.}, 
we find
\begin{equation}
\label{dsr}
\dim_{\Comp} {\S}(\rho) = \sum_i (2i-1)\dim(\rho_i)\ .
\end{equation}
The dimension of $\N(\rho)$ is easily obtained from this result:
the dimension of the $G^{\Comp}$ orbit of $Y(\rho)$ is equal to 
the dimension of $G^{\Comp}$ minus that of the centralizer of $Y(\rho)$,
which is the same as that of the centralizer of $X(\rho)$. Hence
\begin{equation}
\label{nsr}
\dim_{\Comp} \N(\rho) = \dim_{\Comp}G-\dim_{\Comp} {\S}(\rho) \ .
\end{equation}
The dimension of the moduli space \eqref{modsp} can now be computed
as follows.\footnote{We thank P. Slodowy for suggesting this route
to Equation \eqref{dim00}. The same result will be obtained in
the next section using Morse theory arguments.}
Assuming that $\N(\rho_+)\subset \overline{\N(\rho_-)}$, the intersection
$\N(\rho_-) \cap {\S}(\rho_+)$ is non-empty, so that the
sum $\N(\rho_-) + {\S}(\rho_+)$ is well-defined. We then have 
\begin{equation}
\dim_{\Comp} ( \N(\rho_-) \cap {\S}(\rho_+) )
=\dim_{\Comp}\N(\rho_-) + \dim_{\Comp} {\S}(\rho_+)  - 
\dim_{\Comp}  ( \N(\rho_-) + {\S}(\rho_+) ) \ .\nonumber
\end{equation}
Since $\N(\rho_+) + {\S}(\rho_+)$ generates all of ${\cal G}^\Comp$
and $\N(\rho_+)\subset \overline{\N(\rho_-)}$,
the last term in the equation above is equal to 
$\dim_{\Comp}G$. Using \eqref{nsr}, we arrive at
\begin{equation}
\label{dim00}
d(\rho_-,\rho_+)\equiv
\dim_{\Comp}\M(\rho_-,\rho_+)
=\dim_{\Comp}{\S}(\rho_+) - \dim_{\Comp}{\S}(\rho_-) 
\end{equation}
where $\dim_{\Comp}{\S}(\rho_)$ can be computed using \eqref{dsr}.
We have tabulated in Table 1 the Dynkin index and dimensions
of nilpotent orbits and centralizer for $N=6$, together
with some further data to be discussed in the next section.
Table 2 gives the dimension of the moduli space for low
values of $N$, computed using \eqref{dim00}.

\begin{table}
\begin{equation*}
\begin{array}{|c||c|c|c|c|c|c|c|c|c|c|c|c|}
\hline
\rho &  ~\ir{6}~ & \ir{5}{\o}\ir{1} & \ir{4}{\o}\ir{2} & 
\ir{4}{\o}\ir{1}^2&~\ir{3}^2 & \ir{3}{\o}\ir{2}{\o}\ir{1} & \ir{3}{\o}\ir{1}^3 &
~\ir{2}^3~ & \ir{2}^2{\o}\ir{1}^2 & \ir{2}{\o}\ir{1}^4 & ~\ir{1}^6~ \\ 
\hline\hline
n(\rho) &6& 5&4&4&3&3&3&2&2&2&1\\ \hline
\D(\rho) & 35 & 20 & 11&10 &8& 5 & 4 &3& 2 & 1 & 0 \\ \hline\hline
\dim_\Comp \N(\rho) & 30 & 28 & 26 & 24 & 24 & 22 & 18&18&16&10&0 \\ \hline
\dim_\Comp {\S}(\rho) & 6 & 8 & 10 & 12 & 12 & 14 & 18 & 18&20&26&36\\ 
\hline\hline
n_>(\rho) & 48&52&56&60&60&64&72&72&76&88&108\\
n_0(\rho) & 35&34&34&31&32&33&26&27&28&19&0\\
n_<(\rho) & 25&22&18&17&16&11&10&9 &4&1&0\\ \hline\hline
n_>^\Comp(\rho) &73&74&74&77&76&75&82&81&80&89&108\\\hline
\end{array}
\end{equation*} 
\caption{Order of nilpotency $n(\rho)$, Dynkin index ${\cal D}(\rho)$, 
dimension of the nilpotent orbit ${\cal N}(\rho)$ and of 
the shifted centralizer ${\cal S}(\rho)$ for six-dimensional 
representations $\rho$ of $SU(2)$, as in
section \ref{nil}. The last four
rows give the number of positive (resp. zero and negative) eigenvalues
of the Hessian of $W$ at the critical point $\rho$, as in 
section \ref{Morse}.}
\end{table}

\begin{table}[t]
\begin{equation*}
\begin{array}{|c||c|c|}
\hline
\rho_- \backslash \rho_+  & ~\ir{2}~ & ~\ir{1}^2~ \\ \hline\hline
\ir{2} &0 & \Comp^2/\Zint_2 \\ \hline
\ir{1}^2 & & 0 \\ \hline
\end{array}
\qquad
\begin{array}{|c||c|c|c|}
\hline
\rho_- \backslash \rho_+  & ~\ir{3}~ & \ir{2}{\o}\ir{1} & ~\ir{1}^3~ \\ 
\hline\hline
\ir{3} & 0 & \Comp^2/\Zint_3 & 6  \\ \hline
\ir{2}{\o}\ir{1} & & 0 & 4 \\ \hline
\ir{1}^3 & &  & 0 \\ \hline
\end{array}
\end{equation*}

\begin{equation*}
\begin{array}{|c||c|c|c|c|c|}
\hline
\rho_- \backslash \rho_+  & ~\ir{4}~ & \ir{3}{\o}\ir{1} & 
~\ir{2}^2 & \ir{2}{\o}\ir{1}^2 & ~\ir{1}^4~ \\ 
\hline\hline
\ir{4} & 0 & \Comp^2/\Zint_4 & 4 & 6 & 12  \\ \hline
\ir{3}{\o}\ir{1} & & 0 &\Comp^2/\Zint_2 & 4 & 10 \\ \hline
\ir{2}^2 & & & 0 & \Comp^2/\Zint_2 & 8 \\ \hline
\ir{2}{\o}\ir{1}^2 & & & & 0 & 6 \\ \hline
\ir{1}^4 & & & & & 0 \\ \hline
\end{array}
\end{equation*}

\begin{equation*}
\begin{array}{|c||c|c|c|c|c|c|c|}
\hline
\rho_- \backslash \rho_+  & ~\ir{5}~ & \ir{4}{\o}\ir{1} & \ir{3}{\o}\ir{2} & 
\ir{3}{\o}\ir{1}^2&~\ir{2}^2{\o}\ir{1} & \ir{2}{\o}\ir{1}^3 & ~\ir{1}^5~ \\ 
\hline\hline
\ir{5} & 0 & \Comp^2/\Zint_5 & 4 & 6 & 8& 12&20  \\ \hline
\ir{4}{\o}\ir{1} & & 0 & xy^4=z^3 & 4 & 6 & 10 & 18 \\ \hline
\ir{3}{\o}\ir{2} & &  & 0 &  \Comp^2/\Zint_2 & 4 & 8 & 16 \\ \hline
\ir{3}{\o}\ir{1}^2 & &  & & 0 &  \Comp^2/\Zint_2 & 6 & 14 \\ \hline
\ir{2}^2{\o}\ir{1} & & & & & 0 & 4& 12 \\ \hline
\ir{2}{\o}\ir{1}^3 & & & & & & 0 & 8 \\ \hline
\ir{1}^5 & & & & & & & 0 \\ \hline
\end{array}
\end{equation*}
\caption{Moduli spaces for domain walls in $SU(N)$, $N\leq 6$. The
entry denotes the complex dimension of the moduli space, or the space
itself if of dimension 2.}\end{table}

\begin{table}
\begin{equation*}
\begin{array}{|c||c|c|c|c|c|c|c|c|c|c|c|c|}
\hline
\rho_- \backslash \rho_+  & ~\ir{6}~ & \ir{5}{\o}\ir{1} & \ir{4}{\o}\ir{2} & 
\ir{4}{\o}\ir{1}^2&~\ir{3}^2 & \ir{3}{\o}\ir{2}{\o}\ir{1} & \ir{3}{\o}\ir{1}^3 &
~\ir{2}^3~ & \ir{2}^2{\o}\ir{1}^2 & \ir{2}{\o}\ir{1}^4 & ~\ir{1}^6~ \\ 
\hline\hline
\ir{6} & 0 & \Comp^2/\Zint_6 & 4 & 6 & 6& 8& 12&12&14&20&30  \\ \hline
\ir{5}{\o}\ir{1} & & 0 & 2 & 4       & 4& 6& 10&10&12&18&28  \\ \hline
\ir{4}{\o}\ir{2} & &  & 0 &  \Comp^2/\Zint_2 & \Comp^2/\Zint_2 & 4& 8&8&10&16&26  \\ \hline
\ir{4}{\o}\ir{1}^2  & &  & & 0 &  0 & 2& 6&6&8&14&24  \\ \hline
\ir{3}^2  & &  & & 0 &  0 & 2& 6&6&8&14&24  \\ \hline
\ir{3}{\o}\ir{2}{\o}\ir{1} & & & & & & 0 
  &  4 & 4 & 6 & 12 & 22 \\ \hline
\ir{3}{\o}\ir{1}^3 & & & & & & &0 & 0 & 2& 8 & 18 \\ \hline
\ir{2}^3  & & & & & & &0 & 0 & 2& 8 & 18 \\ \hline
\ir{2}^2{\o}\ir{1}^2 & & & & & & & & & 0 & 6& 16 \\ \hline
\ir{2}{\o}\ir{1}^4 & & & & & & & & & & 0 & 10 \\ \hline
\ir{1}^6 & & & & & & & & & & & 0 \\ \hline
\end{array}
\end{equation*} 
\begin{center} {\bf Table 2} ({\it continued}) \end{center}
\end{table}

These tables call for a number of observations. 

(i) All spaces have an even complex dimension, as required
by the hyperk{\"a}hler property. A vanishing dimension means
that the domain wall does not exist, since the translational
zero-mode is always present. 

(i) All spaces have an even complex dimension, as required
by the hyperk{\"a}hler property. A vanishing dimension means
that the domain wall does not exist, since the translational
zero-mode is always present. 

(ii) All allowed domain walls satisfy $W(\rho_-)>W(\rho_+)$
as well as the stronger condition \eqref{cns}, even though this 
criterium has not been used in deriving them. Starting
from $N=6$, we find cases such as
$\ir{4}{\o}\ir{1}^2\to\ir{3}^2$ or $\ir{3}{\o}\ir{1}^3\to\ir{2}^3$
where no BPS domain wall exists, even though the superpotential
does decrease.  Similarly, from $N=8$ on we find pairs
of vacua with same superpotential such as 
$\ir{3}\oplus \ir{1}^5$ and $\ir{2}^4$, for which the same
criterium excludes the existence of tensionless domain walls.
Of course, non-BPS 
domain walls with these boundary conditions do exist, so that 
supersymmetry is spontaneously broken.
on the basis of the superpotential alone.

(iii) A number of spaces have complex dimension $2$ and 
correspond to unresolved ALE spaces;
the case $\ir{4}{\o}\ir{1}\to\ir{3}{\o}\ir{2}$ is particularly interesting
since it exhibits a non-isolated singularity.


(iv) The dimensions of the moduli spaces satisfy the additivity rule
\begin{equation}
\label{addrule}
d(\rho_-,\rho_+)=d(\rho_-,\rho_0)+d(\rho_0,\rho_+)\ ,
\end{equation}
as follows from \eqref{dim00}.
This additivity rule  suggests that
all domain walls can be seen as composite of elementary
domain walls, for which $W(\rho_-)-W(\rho_+)$ is minimal (but strictly
positive) at fixed $\rho_+$: the latter appear above the diagonal
in Tables 2. It is thus sufficient to understand the moduli
of the elementary domain walls. 
To this end, note that besides translations there are two other
classes of continuous deformations of the solutions,
(a) $\Phi^a(t) \to g \Phi^a(t) g^{\dag}$ , where $g$ is a global
transformation in the gauge group $G$;
and
(b) $\Phi^a(t) \to  \rho_+(R) (R^a_{\ b} \Phi^b (t)) \rho_+(R)^\dag$ , with 
$R$ an $SU(2)$ rotation. 
Deformations (a) preserve the boundary conditions \eqref{xbound}
iff  $g$ is in the centralizer of $\rho_+$.
Deformations (b) preserve the boundary conditions for
arbitrary $R$; they act non-trivially
except for a finite subgroup $\Gamma$ of  SU(2), since
(b) acts at $t=-\infty$ as a transformation
(a) with $g = \rho_+(R)\rho_-(R^{-1})$ and  $\rho_+$ and
$\rho_-$ are different representations.
Hence (b) gives an isometric action of $SU(2)$ on the moduli space
with three-dimensional orbits \cite{Kronheimer}. 
The $SU(2)$ action plus translations accounts for the moduli of the
$[n+1]\to [n]\oplus \ir{1}$ walls, as well as all elementary
walls of complex dimension two  in tables 2 (elementary walls correspond
to entries just above the diagonal).
For those cases, the actions of (a) and (b) can be shown to be
equivalent.
The remaining elementary walls have moduli beyond
those of (b) which  can be all understood in terms of (a). 
For example the   $\ir{2}\oplus \ir{1}^{N-2}\to \ir{1}^{N}$
walls (last non-zero entry on the lower right corner of the tables)
have one translation, and $N^2-1$ group conjugations  out of which
those corresponding to the subgroup $U(N-2)$ act trivially. This makes
a total of $4(2N-1)$ real moduli, in agreement with the results of the
tables. Note that transformations (a) and (b) are not independent, as
can be verified easily in some examples.
Hence the elementary walls have no
moduli other than translations, plus global $G$ and R-symmetry
rotations. The `additivity' rule furthermore shows that the moduli of composite
walls can be also accounted for by these operations.

\subsection{Morse theory and dimension of moduli space\label{Morse}}
We now would like to rederive the dimension of the moduli
space $\M(\rho_-,\rho_+)$,  obtained in \eqref{dim00}, in  a different way. 
We will  assuming that the space is
non-empty, and then use  simple arguments about solutions of  
ordinary differential equations. One advantage
 of this more simple-minded approach is
that it shows the robustness of the result under 
small deformations of the K{\"a}hler
potential. 

Assume then that there exists a solution
of a gradient flow in $\Real^N$, 
\begin{equation}
\label{flow}
\frac{d\phi ^i}{dt}= -g^{ij} \frac{\partial W}{\partial\phi ^j}\ ,
\end{equation}
interpolating between two critical points $\rho_-$ and $\rho_+$
of the (real) potential $W$at $\pm \infty$. We now make use
of the following fact: {\it the number of zero modes around solutions
flowing between two different vacua of $W$ is given
generically by the difference of the number of strictly positive
eigenvalues of the Hessian of $W$ at both ends of the flow}.
The reason for this is illustrated on Fig. \ref{flowf}: let us
choose a small sphere $S^{N-1}$ around the point $\rho_-$ at $-\infty$,
and choose the initial condition $\phi$ on that sphere close to the
trajectory we wish to perturb. As we evolve backward in time, 
the solution will not reach $\rho_-$ unless $\phi$ is on the
separatrix between strictly positive and negative eigenvalues 
of the Hessian of $W$ at $\rho_-$: this puts $n_{>}(\rho_-)$ conditions
on $\phi$ on $S^{N-1}$, where $n_{>}(\rho)$ is the number
of strictly positive eigenvalues of the Hessian at the critical
point $\rho$. Now let us evolve the solution forward
in time until we reach the vicinity of $\rho_+$: similarly,
the solution will not attain the critical point at $\rho_+$
unless it arrives along the separatrix, which puts $n_{\leq}(\rho_+)$
conditions on $p$. Altogether, the dimension of the space
of solutions of \eqref{bps} interpolating between $\rho_\pm $ at
$\pm \infty$ and close to the solution of reference
reads\footnote{In line with \eqref{xbound}, we have assumed that the
boundary conditions require $\phi ^i$ to be {\it at} the critical 
point $\rho_+$ at  $+\infty$, and {\it on} the critical locus
$\rho_-$ at $-\infty$. This fixes the zero-modes of the solution
at $+\infty$.}
\begin{equation}
\label{dim0}
d(\rho_-,\rho_+)=(N-1)-n_{>}(\rho_-)-n_{\leq}(\rho_+)+1
=n_{>}(\rho_+)-n_{>}(\rho_-)
\end{equation}
where we added in the translational zero-mode. This is in agreement
with the intuitive 
fact that the number of positive eigenvalues of the Hessian
should increase from one fixed point to another
along the gradient flow \eqref{flow}. 
Note also that the statement in \eqref{dim0} applies
only to generic trajectories, for which the conditions we imposed
at $\pm \infty$ are independent. For degenerate cases, the r.h.s.
of \eqref{dim0} gives only a {\it lower} bound on the dimension
of the moduli space.

\EPSFIGURE[t]{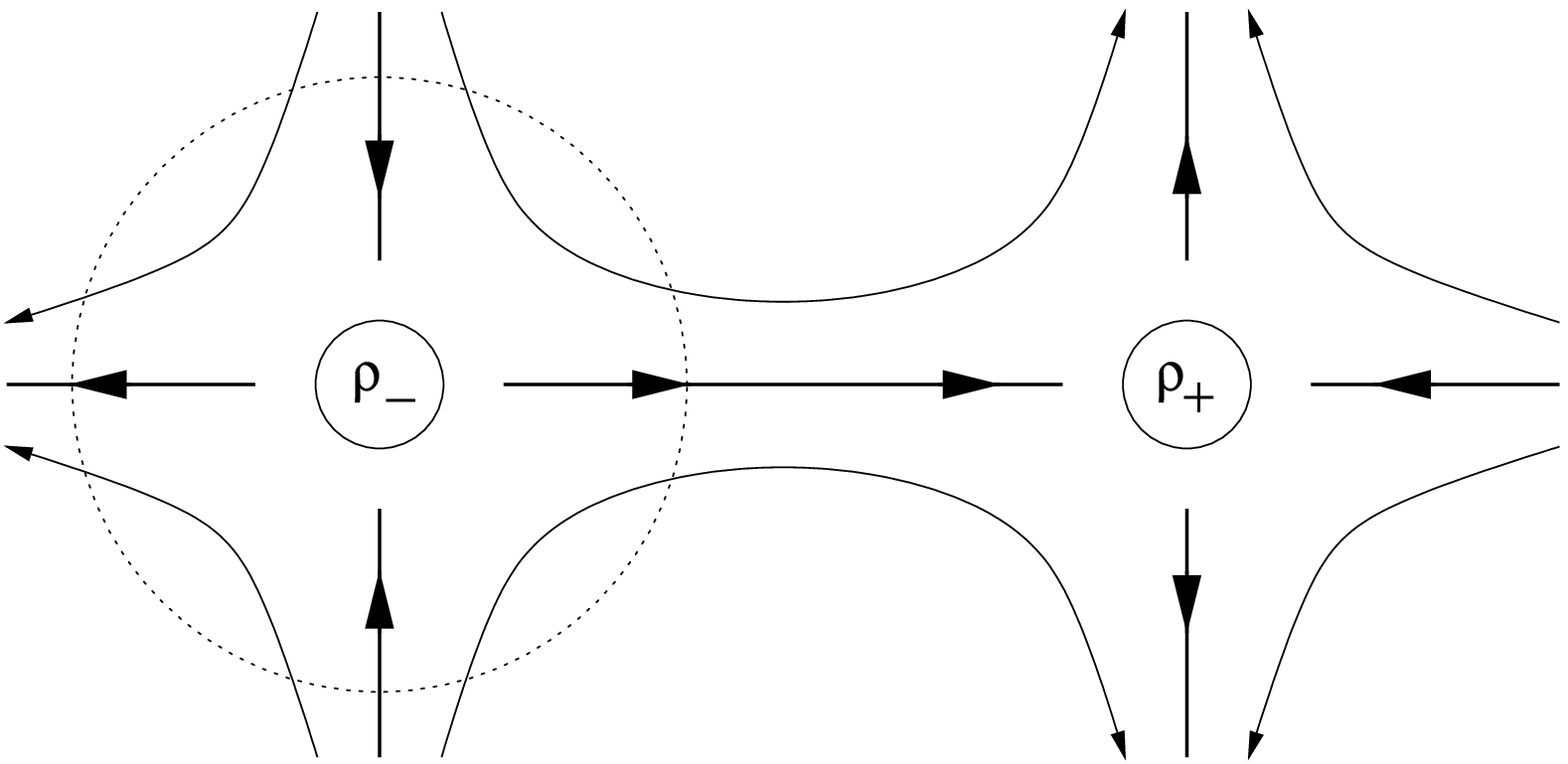,height=6cm}{Gradient flow between two vacua\label{flowf}.}

In order to compute the dimension, we thus need to study the eigenvalues
of the Hessian of $W$ at a critical point \eqref{vacua},  given by
the operator
\begin{equation}
{\cal H}_\rho:\quad K_a \to
 - \epsilon^{abc} [m {\rho}^b,K^c] +  m K_a
\end{equation}
acting on a triplet of antihermitian $N\times N$ matrices $K_a$. 
We note that the spectrum of this
operator is simply obtained from that of 
\begin{equation}
\label{Hrule}
{\cal J}_{\rho}:\quad k^a\to
  \epsilon^{abc} {\rho}^b k^c  \ ,
\end{equation}
through
\begin{equation}
Sp({\cal H}_{\rho}) = m \left[ 1- Sp({\cal J}_{\rho\otimes\rho})  \right]\ .
\end{equation}
${\cal J}_{\rho\otimes\rho}$ acts in the same vector space 
as ${\cal H}_{\rho}$ but now seen as a space of triplet of
$N^2$-dimensional vectors. The spectrum
of ${\cal J}$ is easily computed on irreducible representations,
since ${\cal J}_\rho=J^a_\rho\otimes J^a_{\ir{[3]}}$,
\begin{equation}
\label{Jirrep}
Sp({\cal J}_{\ir{[2j+1]}})= \left\{ (j+1)\vert_{2j-1},\ 
1\vert_{2j+1},\ -j\vert_{2j+3}\right\}\quad (j>0)\ ,\ 
Sp({\cal J}_{\ir{[1]}})=\left\{ 0\vert_3\right\}\ ,
\end{equation}
where we denoted the multiplicity in subscript,
and in arbitrary representations using the sum rule
\begin{equation}
\label{Jrule}
Sp({\cal J}_{\rho\oplus \rho'})=Sp({\cal J}_{\rho}) \cup
Sp({\cal J}_{\rho'})\ .
\end{equation}
We can thus compute the spectrum of ${\cal H}_\rho$ by reducing the
tensor product $\rho\otimes\rho$ into irreductible components and
using \eqref{Jirrep}. In particular, it is easy to see that
the number of zero eigenvalues of ${\cal H}_\rho$ is 
$n_0(\rho)=N^2-\sum n_i^2$,
which is also the number of broken generators
in the vacuum \eqref{vacua}: the only flat directions of the superpotential
are therefore gauge rotations as expected. The number of strictly positive
eigenvalues can also be computed, and is given by
\begin{equation}
\label{nge}
n_{>}(\rho)=N^2+ 2 \sum_i (2i-1)\dim(\rho_i)
\end{equation}
where the $\rho_i$ are the irreducible blocks appearing in the
representation $\rho$, ordered by decreasing dimension as in
\eqref{dsr}. Using \eqref{dim0}, we have thus reproduced the
dimension formula \eqref{dim00}. This confirms that 
the equality in \eqref{dim0}
holds for any choice of ordered representations $\rho_-,\rho_+$.
Furthermore, it shows that the dimension formula is given
by an index, and hence is robust under small deformations
of the K{\"a}hler potential ({\it i.e} the metric $g^{ij}$
appearing in \eqref{flow}).  

The number of strictly negative eigenvalues, or Morse index, 
is also interesting,
since it yields the fermion number of the vacuum $\rho$ in
the supersymmetric quantum mechanics. Its parity
is the same as that of $N-\sum n_i$, so that bosonic vacua are
those with $N$ irreductible components modulo 2.
The number of positive, zero and negative eigenvalues for  $N=6$
are displayed in the lower part of Table 1. 

Before proceeding further  let us comment on the original
problem \eqref{bps}, where the matrices $\Phi^a$ were not
assumed to be antihermitian but general complex matrices.
The complex flow \eqref{bps} is equivalent to a gradient
flow for the real and imaginary parts of the $\Phi ^a$,
with a potential $\Re(W)$. 
Since the potential $\Re(W)$ is  harmonic, the numbers of (strictly) positive
and negative eigenvalues are the same. They  are related to the ones
in the real case as follows:
\begin{equation}
n_>^\Comp(\rho)=n_<^\Comp(\rho)=n_>(\rho)+n_<(\rho)\ ,\quad
n_0^\Comp(\rho)=2 n_0(\rho)\ .
\end{equation}
Assuming that the dimension formula still applies,
we find a dimension smaller than the one for  the real problem,
as one can see  by comparing the first and last rows in Table 1.
Such a  conclusion must be  clearly wrong,  since the solutions of the real
problem are also solutions of the complex problem.
 This signals that the restrictions imposed
at $\pm \infty$ are not independent in the complex case,
and gives us confidence that 
the complex problem does not have any extra  solutions.
We have not been able to prove this statement,  however.



\section{D strings in $AdS_5\times S^5$ and holography}

We return now to the magnetic monopole problem discussed in section 3.
What we will show is that the vacua and domain walls of the
${\cal N}^* =1$ theory, arise also when one studies $N$ D-strings
in the near horizon  geometry of  D3-branes.
A single D-string stretching radially
outwards in  $AdS_5$ is a magnetic source in the fundamental
representation of the dual $SU(n)$ gauge  theory \cite{Aharony:2000ti,Rey:1998ik}. 
Several coincident D-strings in `fuzzy sphere' configurations must
be dual to Wilson-'t Hooft lines in higher representations of $SU(n)$, as
we will here try to argue.

The metric and the  Ramond-Ramond background  in  $AdS_5$ 
read:  
\begin{equation}
ds^2 = L^2~ { dy^2 + dx_\mu dx^\mu \over y^2}\ ,\ \ \ {\rm and}
\ \ \ C_{y0ab}= L^2~ {\epsilon_{abc}x^c\over y^5}\ .
\end{equation}
Here $(y,x^\mu $) are the usual  Poincar{\'e} coordinates,
with $x^\mu = (x^0, x^a)$ parameterizing  the worldvolume of the
background D3 branes. 
We chose  a convenient gauge
for the antisymmetric four-form potential, whose field strength ($H=
dC$) must be 
proportional to the volume form of $AdS_5$. The coordinate $y$ is the
inverse radial distance from the D3-branes, and $y=0$ is the AdS
boundary.

  Consider now $N$ D-strings stretching radially outwards from the D3-branes.
 Their worldvolume theory contains (non-abelian)
Dirac-Born-Infeld (DBI)
and Wess-Zumino (WZ) terms, which combine to give an energy functional
\begin{equation}
E = T_D L^2 \int_0^\infty  {dy\over 2 y^6}~ \Tr \left(
{y^2}{\partial X^a \over \partial y}
- \frac{1}{2}\epsilon^{abc} [X^b,X^c] \right)^2 
+ E_{\rm boundary} + \cdots
\label{Dads}
\end{equation}
We have here used $y$ and $t=x^0$ to parametrize the D-string worldvolume.
The dots  include higher-order terms, which can be neglected
thanks to the supersymmetry of the problem, 
as well as  the  rest
mass  of the stretched strings which is a constant. $E_{\rm boundary}$
is a boundary contribution to the energy which we will determine shortly.
 We have assumed a static
configuration, and have also set the 
worldvolume gauge field and the $S^5$ coordinates
to zero. Finally $T_D$ is the D-string tension.

The various terms in the expression \eqref{Dads} 
can be understood as follows: 
 The gradient square and commutator square
terms are the usual lowest-order contributions from the 
DBI action. The powers of $y$ that accompany them 
can be  guessed from the scale invariance 
($y \to \lambda y ,  \ X^\mu \to \lambda X^\mu $) of the problem.  
The cross term in the expansion of the square would have been a total
derivative in flat space-time. Here it contributes a bulk term
which (after integration by parts) can be recognized as  the 
non-Abelian WZ coupling to the
Ramond-Ramond  four-form proposed by Myers \cite{Myers:1999ps}.
The fact that the energy density is a perfect square
is of course a consequence of the unbroken supersymmetries of the background.
Indeed, one could have used this argument to discover 
the non-Abelian WZ coupling.

The supersymmetric configurations of D-strings are 
given by solutions to  the equations
\begin{equation}
y^2\frac{dX^a}{dy} = \frac12 \epsilon^{abc} [X^b,X^c] \ .
\label{nahmads}
\end{equation}
Furthermore the  boundary conditions that respect scale invariance are
\begin{equation}
\label{adsbound}
X^a(y\to 0) = - y {\rho_{-}^a} + {\rm subleading}\ ,\quad
X^a(y\to \infty) = -y{\rho_+^{a}} + {\rm finite}\ .\quad
\end{equation}
By changing variables to $s  = 1/y$,  we recognize immediately the
mathematical problem analyzed  in the previous sections.
The `vacuum states' of the D-strings correspond to the trivial solutions
\begin{equation}
 X^a = -y \rho^a \ \ {\rm for} \ \ {\rm all}\ \ y\ .
\end{equation} 
They correspond to 
 `fuzzy-sphere' bound states,   described by the SU(2)
representation $\rho$. Domain walls give transitions between these
different vacua, as one moves from the ultraviolet ($s=\infty$) towards
the infrared ($s=0$).

 Because of the $\N=4$ supersymmetry, the mass of the bound states
should  only depend on the  total number,  $N$, of D-strings.
 We may use this argument to
show that 
\begin{equation}
E_{\rm boundary} = 0 \ .
\end{equation}
Indeed, the candidate boundary contribution to the energy is $\sim
W(X)/y^4$, which is 
 the difference between the perfect square terms and  the DBI plus WZ
couplings. If such a  term were really present there would be
a $\rho$-dependent contribution to the D-string mass (proportional to
the ultraviolet cutoff $\delta y$). Since this is forbidden by supersymmetry,   
we  conclude that, contrary to
the situation in flat spacetime,  such a  term is not present 
here. This is analogous to the argument given  for 
fundamental strings  in reference  \cite{Drukker:1999zq}.

Following our discussion in section 3, we expect 
a fuzzy-sphere configuration $\rho$ 
of the D-strings to be   dual to a Wilson-'t Hooft line in a  higher
representation of the gauge group. 
There are two natural candidates for this latter: (a) the 
irreducible representation
$r(\rho)$ of section 3, 
 or (b) the reducible representation corresponding to a  Wilson-'t Hooft
loop 
\begin{equation}
\prod_j {\rm tr}^{n_j}\; (U^{2j+1}) \ , 
\end{equation} 
where  ${\rm tr U}$ corresponds to a single D-string, 
in the fundamental representation of $SU(n)$.
It is unclear to us which of the two interpretations is correct.
 The kinks on the D-string worldvolume should
be, in any case,  dual to  topological twist operators acting 
on the above   Wilson-'t Hooft lines. 
Since $W(\rho_-)>W(\rho_+)$ always, the ultraviolet to infrared flow
tends to effectively increase the number of spherical `fuzzy' D3
branes, 
in accordance with the naive entropy expectation. It would be
interesting to associate the superpotential $W$ to an entropy. We hope to
return to these questions  in some future work.

\acknowledgments
The authors are grateful to V. Bach, 
V.J. Balasubramanian,
S. Cherkis, M. Douglas, P. Fayet, N. Hitchin,
C. Houghton, J. H{\"u}bschmann, B. Kol, J. Maldacena,
W. Nahm, N. Nekrasov, A. Smilga, C. Sonnenschein, 
P. Slodowy, T. Tomaras, D. Tong, 
C. Vafa and  S. T. Yau for useful discussions and remarks.
The work of B. P. is supported in part by the David and Lucile Packard
foundation. C. B. and J. H. would like to thank Harvard University for its
hospitality during the early stages of this work.
Finally, we would like to  acknowledge the support of  
the European  networks FMRX-CT96-0012,
FMRX-CT96-0045, and  HPRN-CT-2000-00122. 
Shortly after our work was released, a paper
appeared by A. Frey \cite{Frey} who addresses the gravity description
of the $\N=1^*$ domain walls. We thank him for pointing out 
a counter-example to the existence conjecture we suggested in the
first version of our work. { The conjecture has been  reformulated and
established in the present version.}


\begin{thebibliography}{99}



\bibitem{Vafa:1994tf}
C.~Vafa and E.~Witten,
``A Strong coupling test of S duality,''
Nucl.\ Phys.\  {\bf B431}, 3 (1994)
[hep-th/9408074].

\bibitem{Donagi:1996cf}
R.~Donagi and E.~Witten,
``Supersymmetric Yang-Mills Theory And Integrable Systems,''
Nucl.\ Phys.\  {\bf B460}, 299 (1996)
[hep-th/9510101].

\bibitem{S} M.~Strassler,   ``Messages for QCD from the Superworld,'' 
Prog.~Theor.~Phys.~Suppl. {\bf 131}, 439 (1998)  [hep-lat/9803009].


\bibitem{Dorey:1999sj}
N.~Dorey,
``An elliptic superpotential for softly broken N = 4 supersymmetric  Yang-Mills theory,''
JHEP {\bf 9907}, 021 (1999)
[hep-th/9906011].

\bibitem{Dorey:2000fc}
N.~Dorey and S.~P.~Kumar,
``Softly-broken N = 4 supersymmetry in the large-N limit,''
JHEP {\bf 0002}, 006 (2000)
[hep-th/0001103].

\bibitem{Girardello:2000bd}
L.~Girardello, M.~Petrini, M.~Porrati and A.~Zaffaroni,
``The supergravity dual of N = 1 super Yang-Mills theory,''
Nucl.\ Phys.\  {\bf B569}, 451 (2000)
[hep-th/9909047].

\bibitem{Polchinski:2000uf}
J.~Polchinski and M.~J.~Strassler,
``The string dual of a confining four-dimensional gauge theory,''
hep-th/0003136.
 
\bibitem{Aharony:2000nt}
O.~Aharony, N.~Dorey and S.~P.~Kumar,
``New modular invariance in the N = 1* theory, operator mixings and 
 supergravity singularities,''
JHEP {\bf 0006}, 026 (2000)
[hep-th/0006008].



\bibitem{Abraham:1991nz}
E.~R.~Abraham and P.~K.~Townsend,
``Intersecting extended objects in supersymmetric field theories,''
Nucl.\ Phys.\  {\bf B351}, 313 (1991).

\bibitem{Cvetic:1991vp}
M.~Cvetic, F.~Quevedo and S.~Rey,
``Stringy domain walls and target space modular invariance,''
Phys.\ Rev.\ Lett.\  {\bf 67}, 1836 (1991).

\bibitem{Cecotti:1993rm}
S.~Cecotti and C.~Vafa,
``On classification of N=2 supersymmetric theories,''
Commun.\ Math.\ Phys.\  {\bf 158}, 569 (1993)
[hep-th/9211097].

\bibitem{Dvali:1997xe}
G.~Dvali and M.~Shifman,
``Domain walls in strongly coupled theories,''
Phys.\ Lett.\  {\bf B396}, 64 (1997)
[hep-th/9612128];
A.~Kovner, M.~Shifman and A.~Smilga,
``Domain walls in supersymmetric Yang-Mills theories,''
Phys.\ Rev.\  {\bf D56}, 7978 (1997)
[hep-th/9706089].


\bibitem{Witten:1997ep}
E.~Witten,
``Branes and the dynamics of {QCD},''
Nucl.\ Phys.\  {\bf B507}, 658 (1997)
[hep-th/9706109].

\bibitem{Gibbons:1999np}
G.~W.~Gibbons and P.~K.~Townsend,
``A Bogomolnyi equation for intersecting domain walls,''
Phys.\ Rev.\ Lett.\  {\bf 83}, 1727 (1999)
[hep-th/9905196].


\bibitem{Nahm} W. Nahm, Phys. Lett. {\bf B90} (1980) 413; and
``The Construction of all Self-Dual Multimonopoles by the ADHM
Method,'' in ``Monopoles in QFT'', Craigie {\it et al} eds.,
WorldScientific (1982).  

\bibitem{hitchin} N.~Hitchin, ``On the Construction of Monopoles''
Com.\ Math.\ Phys. {\bf 89} (1983) 145~.

\bibitem{Ganoulis:1982sx}
N.~Ganoulis, P.~Goddard and D.~Olive,
``Selfdual Monopoles And Toda Molecules,''
Nucl.\ Phys.\  {\bf B205}, 601 (1982).


\bibitem{Kronheimer}
P.~B.~Kronheimer,
``Instantons And The Geometry Of The Nilpotent Variety,''
J.\ Diff.\ Geom.\  {\bf 32} (1990) 473.

\bibitem{Kaplunovsky:1999vt}
V.~S.~Kaplunovsky, J.~Sonnenschein and S.~Yankielowicz,
``Domain walls in supersymmetric Yang-Mills theories,''
Nucl.\ Phys.\  {\bf B552}, 209 (1999)
[hep-th/9811195].



\bibitem{Myers:1999ps}
R.~C.~Myers,
``Dielectric-branes,''
JHEP {\bf 9912}, 022 (1999)
[hep-th/9910053].

\bibitem{Kabat}
D.~Kabat and W.~I.~Taylor,
``Spherical membranes in matrix theory,''
Adv.\ Theor.\ Math.\ Phys.\  {\bf 2}, 181 (1998)
[hep-th/9711078];
S.~Rey,
``Gravitating M(atrix) Q-balls,''
hep-th/9711081.

\bibitem{DBS} 
C.~Bachas, M.~Douglas and C.~Schweigert,
``Flux stabilization of D-branes,''
JHEP {\bf 0005}, 048 (2000)
[hep-th/0003037]~;
J.~Pawelczyk,
``SU(2) WZW D-branes and their noncommutative geometry from DBI action,''
hep-th/0003057.

\bibitem{sus}
J.~McGreevy, L.~Susskind and N.~Toumbas,
``Invasion of the giant gravitons from Anti de Sitter space,''
JHEP {\bf 0006}, 008 (2000)
[hep-th/0003075].


\bibitem{li}
M.~Li,
``Fuzzy gravitons from uncertain spacetime,''
hep-th/0003173~;
P.~Ho and M.~Li,
``Fuzzy spheres in AdS/CFT correspondence and holography from  noncommutativity,''
hep-th/0004072.

\bibitem{Ale}
A.~Y.~Alekseev, A.~Recknagel and V.~Schomerus,
JHEP {\bf 0005}, 010 (2000)
[hep-th/0003187].


\bibitem{Jev}
A.~Jevicki, M.~Mihailescu and S.~Ramgoolam,
``Noncommutative spheres and the AdS/CFT correspondence,''
hep-th/0006239.

\bibitem{Dia}
D.~Diaconescu,
``D-branes, monopoles and Nahm equations,''
Nucl.\ Phys.\  {\bf B503}, 220 (1997)
[hep-th/9608163].


\bibitem{Porrati:1998ej}
M.~Porrati and A.~Rozenberg,
``Bound states at threshold in supersymmetric quantum mechanics,''
Nucl.\ Phys.\  {\bf B515}, 184 (1998)
[hep-th/9708119].

\bibitem{Kac:2000av}
V.~G.~Kac and A.~V.~Smilga,
``Normalized vacuum states in N = 4 supersymmetric 
Yang-Mills quantum  mechanics with any gauge group,''
Nucl.\ Phys.\  {\bf B571}, 515 (2000)
[hep-th/9908096].

\bibitem{Nekrasov:1999cg}
N.~A.~Nekrasov,
``On the size of a graviton,''
hep-th/9909213.

\bibitem{Pouliot}
J.~Polchinski and P.~Pouliot,
``Membrane scattering with M-momentum transfer,''
Phys.\ Rev.\  {\bf D56}, 6601 (1997)
[hep-th/9704029].

\bibitem{Hacquebord}
S.~B.~Giddings, F.~Hacquebord and H.~Verlinde,
``High energy scattering and D-pair creation in matrix string theory,''
Nucl.\ Phys.\  {\bf B537}, 260 (1999)
[hep-th/9804121].

\bibitem{Dynkin}
E.~B.~Dynkin, ``Semisimple subalgebras of semisimple Lie algebras'',
Amer. Math. Soc. Transl. (2) {\bf 6} (1957) 111.

\bibitem{KST}
A.~Kapustin and S.~Sethi,
``The Higgs branch of impurity theories,''
Adv.\ Theor.\ Math.\ Phys.\  {\bf 2}, 571 (1998)
[hep-th/9804027].
D.~Tsimpis,
``Nahm equations and boundary conditions,''
Phys.\ Lett.\  {\bf B433}, 287 (1998)
[hep-th/9804081].

\bibitem{Brecher}
D.~Brecher,
``BPS states of the non-Abelian Born-Infeld action,''
Phys.\ Lett.\  {\bf B442}, 117 (1998)
[hep-th/9804180].


\bibitem{CMT}
N.~R.~Constable, R.~C.~Myers and O.~Tafjord,
``The noncommutative bion core,''
Phys.\ Rev.\  {\bf D61}, 106009 (2000)
[hep-th/9911136].

\bibitem{mald}
C.~G.~Callan and J.~M.~Maldacena,
``Brane dynamics from the Born-Infeld action,''
Nucl.\ Phys.\  {\bf B513}, 198 (1998)
[hep-th/9708147];

\bibitem{Gibbons:1999te}
G.~W.~Gibbons,
``Branes as BIons,''
Class.\ Quant.\ Grav.\  {\bf 16}, 1471 (1999)
[hep-th/9803203].

\bibitem{Hurtubise}
J.~Hurtubise and M.~K.~Murray,
``On The Construction Of Monopoles For The Classical Groups,''
Commun.\ Math.\ Phys.\  {\bf 122} (1989) 35.

\bibitem{Cherkis}
S.~A.~Cherkis and A.~Kapustin,
``Singular monopoles and gravitational instantons,''
Commun.\ Math.\ Phys.\  {\bf 203}, 713 (1999)
[hep-th/9803160];
S.~A.~Cherkis and A.~Kapustin,
``D(k) gravitational instantons and Nahm equations,''
Adv.\ Theor.\ Math.\ Phys.\  {\bf 2}, 1287 (1999)
[hep-th/9803112];
S.~A.~Cherkis and A.~Kapustin,
``Singular monopoles and supersymmetric gauge theories in three  dimensions,''
Nucl.\ Phys.\  {\bf B525}, 215 (1998)
[hep-th/9711145].

\bibitem{Weinberg}
E.~J.~Weinberg,
``Duality and Massless Monopoles,''
hep-th/0007038.




\bibitem{Donaldson}
S.~K.~Donaldson,
``Nahm's Equations And The Classification Of Monopoles,''
Commun.\ Math.\ Phys.\  {\bf 96} (1984) 387.


\bibitem{Slodowy}
P. Slodowy, ``Simple singularities and simple algebraic groups'',
Lecture Notes in Math. {\bf 815}, Springer, Berlin, 1980.


\bibitem{Kron2}
P.~B.~Kronheimer,
``The Construction Of ALE Spaces As Hyperkahler Quotients,''
J.\ Diff.\ Geom.\  {\bf 29} (1989) 665.


\bibitem{Aharony:2000ti}
O.~Aharony, S.~S.~Gubser, J.~Maldacena, H.~Ooguri and Y.~Oz,
``Large N field theories, string theory and gravity,''
Phys.\ Rept.\  {\bf 323}, 183 (2000)
[hep-th/9905111].

\bibitem{Rey:1998ik}
S.~Rey and J.~Yee,
``Macroscopic strings as heavy quarks in large N gauge theory and  anti-de Sitter supergravity,''
hep-th/9803001.


\bibitem{Drukker:1999zq}
N.~Drukker, D.~J.~Gross and H.~Ooguri,
``Wilson loops and minimal surfaces,''
Phys.\ Rev.\  {\bf D60}, 125006 (1999)
[hep-th/9904191].




\bibitem{Frey}
A.~Frey,
``Brane Configurations of BPS Domain Walls for the $\N=1^*$ $SU(N)$
 Gauge Theory,'', 
JHEP {\bf 0012}, 020 (2000)
[hep-th/0007125].

\end{thebibliography}
\end{document}